\documentclass[useAMS,usenatbib]{mn2e}
\usepackage{times}
\usepackage{graphicx}

\newcommand{\dif}[1]{\ensuremath{\mathrm{d}#1}}

\title[Broad lines for negatively spinning BHs]{Broad emission lines for a negatively spinning black hole}
\author[T. Dauser et al.]{T.\ Dauser$^{1}$\thanks{E-mail:
  thomas.dauser@sternwarte.uni-erlangen.de}, 
  J. Wilms$^{1}$,
  C.S.\ Reynolds$^{2}$,
  and L.W.\ Brenneman$^{3}$\\ 
  $^{1}$ Dr.\ Karl Remeis-Observatory and Erlangen Centre for
  Astroparticle Physics, Sternwartstr.~7, 96049 Bamberg, Germany\\
  $^{2}$ Department of Astronomy  and Maryland Astronomy Center for
  Theory and Computation, University of Maryland, College Park, MD
  20742, USA\\
  $^{3}$ Harvard-Smithsonian Center for Astrophysics, 60 Garden
  Street, Cambridge, MA 02138, USA  
}
\begin{document}

\pagerange{\pageref{firstpage}--\pageref{lastpage}} \pubyear{2010}

\maketitle

\label{firstpage}

\begin{abstract}
  We present an extended scheme for the calculation of the profiles of
  emission lines from accretion discs around rotating black holes. The
  scheme includes discs with angular momenta which are parallel and
  antiparallel with respect to the black hole's angular momentum, as
  both configurations are assumed to be stable. We discuss line shapes
  for such discs and present a code for modelling observational data
  with this scheme in X-ray data analysis programs.  Based on a
  Green's function approach, an arbitrary radius dependence of the
  disc emissivity and arbitrary limb darkening laws can be easily
  taken into account, while the amount of precomputed data is
  significantly reduced with respect to other available models.
\end{abstract}

\begin{keywords}
Accretion, Accretion Discs, black hole physics, Galaxies: Nuclei,
galaxies: active, Lines: Profiles
\end{keywords}

\section{Introduction}

Skew-symmetric, broadened Fe K$\alpha$ emission lines are seen in many
Active Galactic Nuclei (AGN) such as MCG$-$6-30-15
\citep{tanaka1995a,wilms2001a,Miniutti2007amnras}, 1H0707$-$495
\citep{fabian2009a}, and others \citep{nandra2007a}, Galactic black
hole binaries such as Cygnus~X-1 \citep{Fabian1989,miller2002a},
GX~339$-$4 \citep{miller2004a,caballerogarcia2009a}, or GRS~1915+105
\citep{Martocchia2002a,blum2009a}, and neutron star systems
\citep{disalvo2009a,cackett2008a,cackett2009a,shaposhnikov2009a}.
These lines are generally interpreted as being caused by the
relativistic motion of the line emitting material close to the central
compact object. Since the line shape depends on the spin of the black
hole, $a$, and the emissivity and inclination of the surrounding
accretion disc, the diagnostic power of relativistic lines is very
high, as they provide one of the most direct ways to probe the physics
of the region of strong gravity close to the black hole \citep[see,
e.g., ][for a review]{reynolds2003a}. High signal-to-noise
observations of AGN and Galactic black holes have already resulted in
several measurements of $a$ with formally small error bars
\citep{Brenneman2006a,miller2004a,miller2008a,miller2009a}, with
systematic effects due to the high count rate of Galactic sources
\citep{yamada2009a,done2010a} and due to the uncertainty of the
parameters of the underlying continuum
\citep[e.g.,][]{ross2007a,reynolds2008a} currently dominating the
uncertainty of the measurements.

Observations of AGN in the \textsl{XMM-Newton} and \textsl{Chandra}
deep fields prove that broadened iron lines already occured at high
redshifts, $z$, \citep[but see
\citealt{corral:08a}]{comastri:04a,streblyanska2005a}.
Although recent studies seem to exclude that these broad lines
  are a common feature \citep{Longinotti2008a}, observations of such
lines could therefore be used to study the expected evolution of black
hole spin with $z$. For example, strong changes in amplitude and
direction for the central black hole are predicted in stochastic
evolution models \citep{king2008a,Volonteri2005a}. Observations of
cavities in nearby galaxy clusters are also evidence for spin
evolution \citep{wise2007a,fabian2000a}. In galactic binary systems,
the initial kick during the formation of a stellar-mass black hole in
a supernova can lead to a strong misalignment between the disc and the
black hole \citep{brandt1995a}.

Depending on the mode of accretion, in all of these scenarios it is
possible that the angular momenta of black hole and accretion disc
become antiparallel, i.e., the black hole has ``negative spin''. As
shown by \citet{King2005a}, both parallel and antiparallel alignments
of the disc and black hole angular momenta are stable configurations;
misaligned discs will evolve to one of them. It is therefore not
unlikely that a configuration with antiparallel spins exists in
nature. In fact, accretion onto rapidly-spinning retrograde black
holes may be of some importance for understanding the properties of
powerful radio-loud AGN. Employing the flux-trapping model of
\citet{Reynolds2006a}, \citet{Garofalo2009a} argues that an accretion
disk around a retrograde black hole is a particularly potent
configuration for generating powerful jets. Moreover this
  might also explain the lack of radio-loud AGN in observations
  \citep{Garofalo2010a}. It is tantalizing that the broad iron line
in the powerful radio-loud AGN 3C120 implies a truncation at $r\sim
10\,GM/c^2$ \citep{Kataoka2007a}, very close to the innermost stable
circular orbit (ISCO) for a rapidly-rotating retrograde black hole.
However, a further exploration of this line of thought requires
fully-relativistic iron line models that are valid for retrograde
black holes. Although first calculations of line profiles for
  a negatively spinning black hole were already performed, e.g., by
  \citet{Jaroszynski1997a} and \citet{Schnittmann2006a}, none of the
  currently available models for relativistic lines such as
  \texttt{diskline} \citep{Fabian1989}, \texttt{laor}
  \citep{Laor1991}, \texttt{kerrdisk} \citep{Brenneman2006a}, or the
  \texttt{ky}-family of models \citep{Dovciak2004a} are valid for
  black holes with retrograde accretion discs.

In this \textsl{Paper} we therefore extend the formalism of
\citet{Cunningham1975} employed by many of these models to the case of
$-0.998\le a \le +0.998$. Section~\ref{sec:theory} presents an
overview of the scheme used for the calculations, including a new
approach which reduces significantly the amount of data to be
precalculated and allows for a flexibility in terms of the emissivity
of the accretion disc and the limb-darkening law
(Sect.~\ref{sec:calculations}). The implementation of this scheme in a
new code for calculating relativistic lines, \texttt{relline}, and the
comparison of this scheme with other models is described in
Sect.~\ref{sec:relline}.  Section~\ref{sec:discussion} presents
results for line profiles and summarizes our results.

\section{Theory: Relativistically Broadened Lines} \label{sec:theory}
\subsection{The equations of motion around a rotating black hole} 
\label{sec:motion}
In order to account for the strongly curved space and allow a spinning
black hole, a fully relativistic approach in the \citet{Kerr1963}
metric was chosen. This metric is characterized by the mass, $M$, and
the angular momentum, $J$, of the black hole, which is commonly
parametrized as $a=J/M$.  We will call the black hole in a
  system where it spins in the opposite direction of the accretion
  disc a \emph{negatively spinning} black hole (i.e. $a < 0$).
Throughout this paper, units of $ \mathrm{G} \equiv \mathrm{c} \equiv
1$ are chosen. The line element in \citet{Boyer1967} coordinates is
\begin{eqnarray} \label{eq:kerr_metric}
\dif{s}^2 = & - \left( 1- \frac{2Mr}{\Sigma} \right) \dif{t}^2 -
\frac{4aMr \sin^2{\theta}}{\Sigma} \dif{t} \dif{\varphi}
+{\Sigma}{\Delta} \dif{r}^2 \nonumber \\ & + \Sigma \dif{\theta}^2 +
\left( r^2 + a^2 + \frac{2a^2Mr \sin^2\theta}{\Sigma} \right)
\sin^2\theta \dif{\varphi}^2 \quad ,
\end{eqnarray}
where $\Delta = r^2 - 2Mr +a^2$ and $\Sigma = r^2 +
a^2\cos^2{\theta}$, the angle $\phi$ is measured in the plane of the
disc, and the black hole's angular momentum points towards $\theta=0$.
Taking into account the black hole's interaction with thermal
  photons from the accretion disc, its spin is restricted to $a \le
  a_\mathrm{max} < 1$ as capturing photons with negative angular
  momentum (with respect to the movement of the disc) becomes more
  likely for increasing $a$ and thus prevents a spin up to the extreme
  value of $a=1$ \citep{1974Thorne}. Assuming that a negatively
  spinning system is created by flipping the spin of a system with
  $a>0$ sets the lower limit of the spin at $a \ge -a_\mathrm{max}$,
  as infalling matter from the couterrotating disc clearly decreases
  the absolute value of the spin with time. We choose $a_\mathrm{max}
  = 0.998$, which is commonly used and has been calculated by
  \citet{1974Thorne}.

As we are interested in particle orbits around the black hole, we need
to derive the equations of motion for a test particle in the Kerr
metric. This can, e.g., be done by solving the Geodesic equation
directly, which formally is a general equation of motion for all
possible metrics in General Relativity. Using the conserved quantities
of motion \citep{Carter1968}, i.e., the energy $E$, the angular
momentum $L$, the rest mass $\mu$ of the particle, and
\begin{equation}
\mathcal{Q} = p_\theta^2 + \cos^2\theta\left[a^2(\mu^2 - p_t^2) +
  p_\varphi^2/\sin^2\theta\right] \quad.
\end{equation}
the general equations of motion  are \citep{Bardeen1972}:
\begin{equation}\label{eq:m1}
  \Sigma \dot{t} = -a(aE\sin^2\theta - L) + (r^2 +
  a^2)\frac{T}{\Delta}
\end{equation}
\begin{equation}\label{eq:m2}
\Sigma \dot{r} = \pm \sqrt{V_r}
\end{equation}
\begin{equation}\label{eq:m3}
\Sigma \dot{\theta} = \pm \sqrt{V_\theta} 
\end{equation}
\begin{equation}\label{eq:m4}
\Sigma \dot{\varphi} = - \left(aE - \frac{L}{\sin^2\theta}
  \right) + a\frac{T}{\Delta} \quad,
\end{equation}
where
\begin{equation}
V_r= T^2 - \Delta \left(\mu^2+r^2 + (L - aE)^2 +
\mathcal{Q} \right) ,
\end{equation}
\begin{equation}
V_\theta = \mathcal{Q} - \cos^2\theta
\left( L^2/\sin^2\theta + a^2(\mu^2 - E^2)\right)
\end{equation}
and
\begin{equation}
T = E(r^2+a^2) -aL \quad. 
\end{equation}
  The signs in Eq.~\ref{eq:m2} and Eq.~\ref{eq:m3} can be chosen
  independently and account for the direction of the photon. The upper
  sign means a movement with growing $r$/$\theta$ and the lower sign
  stands for the opposite behaviour, respectivly. Thus they can be
  chosen arbitrarily, but change, e.g., when a turning point occurs.

\subsection{The accretion disc}
For simplicity we assume a geometrically thin accretion disc which
lies in the equatorial plane of the system, i.e., $\theta = \pi/2$ and
$\dot{\theta} = 0$. Additionally we require the disc to be stationary
and to consist of particles orbiting the compact object on circular
orbits. This approach fully determines the system \citep{Bardeen1972}.
Taking into account that the particles can be on pro- and retrograde
orbits with respect to the spinning direction of the black hole, the
particles have an angular velocity
\begin{equation}\label{eq:omega}
 \omega =  \frac{\sqrt{M}}{r\sqrt{r} + a \sqrt{M}} \quad.
\end{equation}
The four-velocity of the accretion disc is given by
\begin{equation}
u^\mu = u^t(\partial_t + \omega \partial_\phi)
\end{equation}
where
\begin{equation}\label{eq:four_velocity}
u^t = \frac{r\sqrt{r} + a\sqrt{M}}{\sqrt{r}\sqrt{r^2 - 3Mr
   + 2a\sqrt{M}\sqrt{r}}} \quad.
\end{equation}
Further calculation reveals \citep{Bardeen1972} that there exists a
radius of \emph{marginal stability} (often referred to as innermost
stable circular orbit, ISCO) at
\begin{equation}
  r_\mathrm{ms}(a) = M\left( 3 + Z_2 - \mathrm{sgn}(a)
  \sqrt{(3-Z_1)(3+Z_1+2Z_2)}\right) \quad,
\end{equation}
where 
\begin{equation}
Z_1 = 1 + ( 1- a^2)^{1/3}\left[(1+a)^{1/3} + (1-a)^{1/3}
  \right]
\end{equation}
and
\begin{equation}
Z_2 = \sqrt{3a^2 + Z_1^2}\quad.
\end{equation}
Thus the inner edge of the accretion disc has to be at a radius
$r_\mathrm{in}>r_\mathrm{ms}$, as no stable circular orbits can exist
inside of it. The minimum inner radius of an accretion disc
is $r_{\mathrm{ms}}(a=+0.998)=1.23\,r_\mathrm{g}$, where $r_\mathrm{g}
= GM/c^2$ is called gravitational radius. This orbit is only possible
for particles circulating around a maximally rotating black hole with
positive angular momentum, as the orbits are supported by
frame-dragging effects. In the case of a negative spin, the same
effects push the inner edge out to $r_{\mathrm{ms}}(a=-0.998) \sim 9
\,r_\mathrm{g}$.

\subsection{Radiation transport}
Having described the accretion disc, i.e. the frame where the
  photons are emitted, we can now trace them back to a distant
observer. Following, e.g., \citet{Krolik1999}, we use the effective
Lagrangian
\begin{equation}
\mathcal{L}_\mathrm{eff} = \frac{1}{2}g_{\mu\nu}\dot{x}^\mu \dot{x}^\nu
\end{equation}
for massless particles ($\mu = 0$) in order to calculate the photon's
momentum
\begin{equation} \label{eq:photon_momentum}
  p_\mu = \dif{\mathcal{L}}/\dif{x^\mu} = 
  E \left( -1,\: \pm 1/\Delta\sqrt{V_r},\:
  \pm  \sqrt{V_\theta},\:
  \lambda \right)^\mathrm{T} \quad.
\end{equation}
where $E$ is the energy of the photon in flat space and thus in very
good approximation the photon's energy measured by an observer at
large distance. $\lambda = L/E^2$ can be interpreted as the angular
momentum.

As the photons originate from a stationary and axi-symmetric accretion
disc, we only have to consider the motion in $(r,\theta)$ direction.
Using the equations of motion (Eq.~\ref{eq:m1}~-~\ref{eq:m4}), this
assumption reduces the problem of determining the trace of a photon to
solving
\begin{equation} \label{eq:int_motion}
 \int\limits_{r_\mathrm{e}}^{\infty}\frac{\dif{r}}{\sqrt{V_r}} =
 \int\limits_{\pi/2}^{\theta_\mathrm{o}}\frac{\dif{\theta}}{\sqrt{V_\theta}}
\end{equation}
where the integration is carried out from the point of emission
$(r,\theta) = (r_\mathrm{e},\pi/2)$ to the observer at
$(\infty,\theta_\mathrm{o})$ \citep{Carter1968}. Here $r_\mathrm{e}$
is the radius of emission and $\theta_\mathrm{o}$ the angle with
respect to the normal of the accretion disc at which the photon is
observed. Note that solving Eq.~\ref{eq:int_motion} fully determines
all constants of motion and thus the momentum and therefore the
emission angle of the photon, $\theta_\mathrm{e}$, is fixed. By
splitting the integral into parts and keeping track of the correct
sign in front of $\sqrt{V_r}$ and $\sqrt{V_\theta}$ (see
Sec.~\ref{sec:motion}), turning points in $r$ and $\theta$ direction
can be taken into account.

\begin{figure}
    \includegraphics[width=84mm]{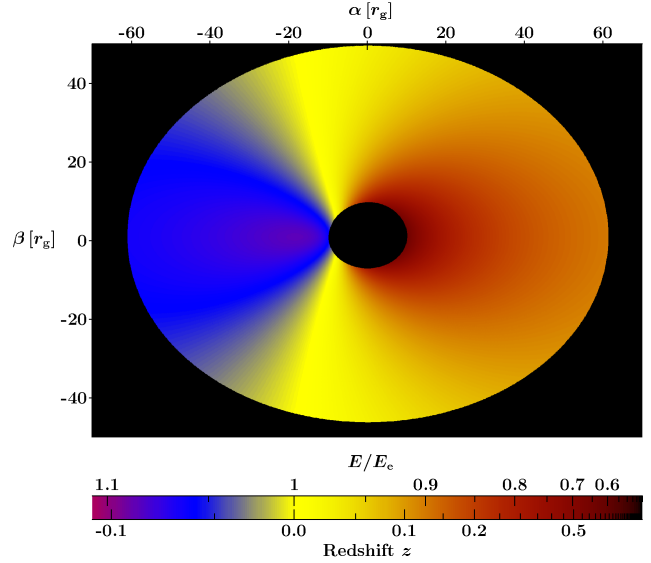}
    \caption{ \label{fig:disc} Map of a retrograde accretion
      disc around a maximally spinning black hole ($a=-0.998$) as seen
      from a distant observer at an inclination angle of
      $\theta_\mathrm{o} = 40^\circ$. The disc truncates at
      $60\,r_\mathrm{g}$. $\alpha$ and $\beta$ are the coordinates
      defined on the plane of the sky (i.e., perpendicular to the line
      of sight; see Eq.~\ref{eq:alpha_beta}). The color code shows the
      energy shift of the photons, asymmetries are due to relativistic
      light bending. The blue-shifted left part of the disc moves
      towards the observer, whereas the right part recedes from the
      observer.}
\end{figure}

To calculate the specific intensity $I_E^{\mathrm{obs}}
(\theta_\mathrm{o})$ observed under the angle $\theta_\mathrm{o}$ at
energy $E$, we have to integrate over the local specific intensity
$I_{E_\mathrm{e}}(r_\mathrm{e},\theta_\mathrm{e})$ emitted from the
accretion disc. As the direction of a photon changes along its way to the
observer due to strong gravity, as shown by \citet{Cunningham1973}
this integration is easily performed after the disc is projected onto
a plane perpendicular to the line of sight, spanned by the impact
parameters $\alpha$ and $\beta$, which are connected to the solid
angle through \citep{Cunningham1973}
\begin{equation}\label{eq:alpha_beta}
 \dif{\alpha}\,\dif{\beta} = D^2 \dif{\Omega} \quad,
\end{equation}
where $D$ is the distance to the observer. In order to obtain the
specific intensity on the projected plane, we use Liouville's theorem,
$I_E/E^3 = \mathrm{const.}$ \citep{Lindquist1966a}. Integrating over
the projected accretion disc then yields
\begin{equation}\label{eq:lspec0}
  I_E^{\mathrm{obs}} (\theta_\mathrm{o})  =  
\int \left(   \frac{E}{E_\mathrm{e}} \right)^3
  I_{E_\mathrm{e}}(r_\mathrm{e},\theta_\mathrm{e})
\,\dif{\alpha}\,\dif{\beta} \quad.
\end{equation}
We define the general relativistic Doppler shift $g=E/E_\mathrm{e}$,
which can be calculated from the four-velocity of the accretion
disc (Eq.~\ref{eq:four_velocity}) and the momentum of the photon
(Eq.~\ref{eq:photon_momentum}) \citep{Cunningham1975},
\begin{equation} \label{eq:redshift}
  g = \frac{E}{E_\mathrm{e}} = -\frac{p^t}{p_\mathrm{e}^\mu u_\mu} =
  \frac{\sqrt{{r_\mathrm{e}}}\sqrt{{r_\mathrm{e}}^2-3M{r_\mathrm{e}}
      + 2a\sqrt{M {r_\mathrm{e}}}} }
       {{r_\mathrm{e}}\sqrt{{r_\mathrm{e}}} + a\sqrt{M} -
         \sqrt{M}\lambda} \;.
\end{equation}
Note that for the special case of viewing the accretion disc from top
($\theta_\mathrm{o}$=0, leading to $\lambda=0$), the purely
gravitational redshift for a negative spin is slightly higher than for
positive $a$. Figure~\ref{fig:disc} shows the redshift of a complete
accretion disc as seen by a distant observer, illustrating the effects
of gravitational redshifting, Doppler boosting, and light bending.

Using the maximum and minimum energy ratio $g$ on a certain ring of
the accretion disc, following \citet{Cunningham1975} we can
additionally define 
\begin{equation}
g^* = \frac{g-g_\mathrm{min}}{g_\mathrm{max} - g_\mathrm{min}} \in [0,1]
\end{equation}
This approach allows us to perform a coordinate transformation from
$(\alpha, \beta)$ to $({r_\mathrm{e}}, g^*)$ and to carry out the integration
(Eq.~\ref{eq:lspec0}) over the accretion disc. The expression for the
coordinate transform is commonly simplified further by introducing the
\emph{transfer function} \citep{Cunningham1975}
\begin{equation}\label{eq:f}
   f(g^*,{r_\mathrm{e}},\theta_\mathrm{o}) = \frac{1}{\pi
     {r_\mathrm{e}}} g\sqrt{g^*(1-g^*)} \left|\frac{\partial (\alpha,
     \beta)}{\partial (g^*, {r_\mathrm{e}})} \right| \;.
\end{equation}
Note that here the determinant can not be calculated analytically and
thus the transfer function has to be evaluated numerically. Using the
above equations, the observed intensity under an inclination angle
$\theta_\mathrm{o}$ then is
\begin{equation}\label{eq:lspec_general}
  I_{E}^{\mathrm{obs}} (\theta_\mathrm{o}) =
  \int\limits_{r_\mathrm{in}}^{r_\mathrm{out}} \int\limits_0^1
  \frac{\pi {r_\mathrm{e}} g^2f(g^*,{r_\mathrm{e}},\theta_\mathrm{e})
  }{\sqrt{g^*(1-g^*)}}
  I_{E_\mathrm{e}}({r_\mathrm{e}},\theta_\mathrm{e})\,\dif{r}\,\dif{g}^*,
\end{equation}
where the inner and outer radii of the accretion disk are
$r_\mathrm{in}$ and $r_\mathrm{out}$, respectively.

\subsection{Calculating Relativistic Line Profiles}
 \label{sec:calculations} 

 The effort to calculate line profiles is almost fully buried in the
 determination of the transfer function, which depends in general on
 four parameters ($a$, $\theta_\mathrm{o}$, ${r_\mathrm{e}}$, and
 $g$). In order to allow for a real-time fitting of observational data
 it is generally necessary to precalculate $f$ or some variant of it.
 The quality of a given model then depends strongly on the amount of
 precalculated information. Table sizes can amount up to several
 hundreds of megabytes.

In most available models, the approach chosen is to precalculate the
value of the inner integral in Eq.~\ref{eq:lspec_general} using a
Gaussian line shape for $I_{E_\mathrm{e}}$ and some prescription of
the limb-darkening law, i.e., the dependence of $I_{E_\mathrm{e}}$
from $\theta_{\mathrm{e}}$. A disadvantage of this approach is that
any change of the limb-darkening law necessitates a recomputation of
the precalculated tables. In order to avoid this problem, we use a
Green's function approach to model the specific intensity originating
from the disc as purely mono-energetic at $E_0$,
\begin{equation}\label{eq:ispec}
  I_{E_\mathrm{e}}({r_\mathrm{e}},\theta_\mathrm{e}) = \delta(E_\mathrm{e}-E_0)
  \varepsilon({r_\mathrm{e}},\theta_\mathrm{e}) \quad,
\end{equation}
The dependencies of the local intensity on the emission angle
$\theta_\mathrm{e}$ (e.g., limb darkening effects) and the radius
${r_\mathrm{e}}$ are described by
$\varepsilon({r_\mathrm{e}},\theta_\mathrm{e})$. Inserting this into
Eq.~\ref{eq:lspec_general} and evaluating the delta function then
gives
\begin{equation}\label{eq:lspec}
I_{E_\mathrm{o}}^{\mathrm{obs}} (\theta_\mathrm{o})
  =\int\limits_{r_\mathrm{in}}^{r_\mathrm{out}} 
  \frac{ \pi g^3{r_\mathrm{e}} f(g^*,{r_\mathrm{e}},\theta_\mathrm{o})}
{E_0(g_\mathrm{max} - g_\mathrm{in})\sqrt{g^*(1-g^*)}}
  \varepsilon({r_\mathrm{e}},\theta_\mathrm{e})\,\dif{r}_\mathrm{e}
\end{equation}
where the transfer function $f$ can be calculated using the code of
\citet{Speith1995}.

As the integration over $r$ is trivial, we focus on the emission of
one radius and its contribution $I^{\mathrm{obs}}_{E_\mathrm{i}}(r)$ to a
certain energy bin $i$ ranging from $E_\mathrm{lo}$ to
$E_\mathrm{hi}$. Using Eq.~\ref{eq:lspec}, 
\begin{equation} \label{eq:lbin}
  I^{\mathrm{obs}}_{E_i}(r) \propto
  \int\limits_{E_\mathrm{lo}}^{E_\mathrm{hi}} 
  \frac{g^3 f(g^*)}{(g_\mathrm{max} -
    g_\mathrm{min})\sqrt{g^*(1-g^*)}} \, \dif{E}   \quad,
\end{equation}
where the integrand diverges at $g^*=0$ and $g^*=1$, i.e., at the two
points on the ring where the minimum and maximum energy shifts occur
with respect to $E_0$. As the divergences imply that these points
contribute significantly to the overall luminosity, great care has to
be taken by numerical integration. In order to avoid numerical
instabilities, we make use of the fact that the dependence of $f(g^*)$
close to these points can be calculated analytically as
\begin{equation}
  f(g^* \rightarrow 0) \propto \sqrt{1-g^*}
  \quad \mathrm{and} \quad
  f(g^* \rightarrow 1) \propto \sqrt{g^*} \quad.
\end{equation}
This leaves the integrand of Eq.~\ref{eq:lbin} with a divergence of
the kind $1/\sqrt{x}$ for $x \rightarrow 0$. Assuming that $g \sim
\mathrm{const.}$ in this energy bin, the integration can be performed
analytically, leading to
\begin{equation}\label{eq:approx}
  I^{\mathrm{obs}}_{E_i}(r) \propto 2(\sqrt{E_\mathrm{hi}} -
  \sqrt{E_\mathrm{lo}}) \quad.
\end{equation}
As the above assumption might not be valid for the whole bin, we
define a criteria by choosing a sufficiently small value of $h$ such
that it is legitimate to use Eq.~\ref{eq:approx} for $g^* \in [0,h]$
and $g^* \in [1-h,1]$. The normalization factors are then determined
from $I^{\mathrm{obs}}_h(r)$ and $I^{\mathrm{obs}}_{1-h}(r)$. For $g^*
\in [h,1-h]$, an adaptive Romberg method is chosen to solve
Eq.~\ref{eq:lspec} directly. These numerical improvements serve to
avoid the spikes seen in some other models and keep the advantages of
the Green's function approach (see below).

\subsection{The \texttt{relline} Model}\label{sec:relline}
\begin{figure}
  \includegraphics[width=84mm]{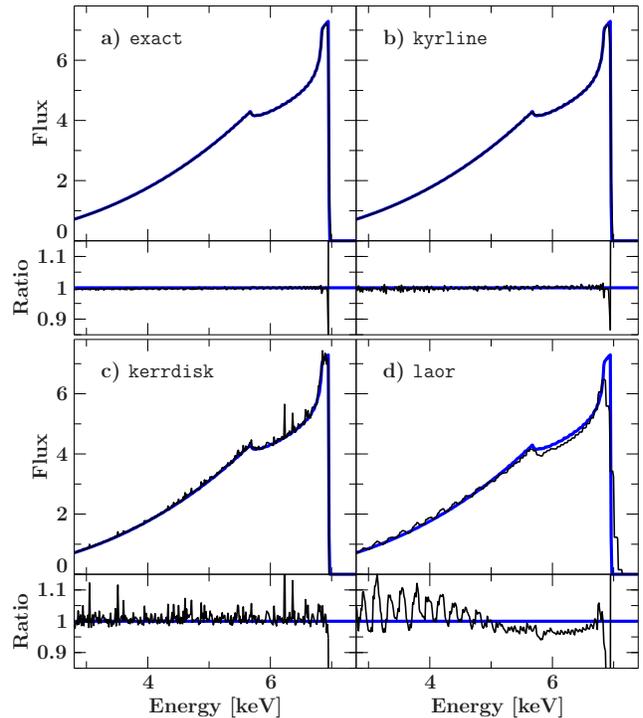}
  \caption{ \label{fig:line_models} Comparison of the \texttt{relline}
    model (blue) with a) the exact line profile, b) \texttt{kyrline}-,
    c) \texttt{kerrdisk}- and d) the \texttt{laor}-model for an
    emission line at $E_0 = 6.4\,\mathrm{keV}$ around a black hole
    with $a=+0.998$, an inclination angle of $\theta_\mathrm{o} =
    40^\circ$, an emissivity $r^{-3}$, and an outer radius of
    $50\,r_\mathrm{g}$. }
\end{figure}

We have implemented the formalism of Sect.~\ref{sec:calculations} as a
model function, called \texttt{relline}, that can be added to data
analysis software such as \textsc{isis} \citep{Houck2000a} or
\textsc{xspec} \citep{Arnaud1996a}. In order to reduce the required
CPU time, following \citet{Brenneman2006a} we calculate the transfer
function (Eq.~\ref{eq:f}) for various combinations of $-0.998 \le a
\le 0.998$ and $0^\circ\le \theta \le 89^\circ$. The highly resolved
line profile is then calculated using a linear interpolation of the
slowly varying transfer function followed by a numerical integration
over $r$ for the returned intensity. The \texttt{relline} model is
provided at
{\scriptsize\texttt{www.sternwarte.uni-erlangen.de/research/relline/}}.
Both an additive and a convolution model are provided (the latter for
calculating the relativistic smearing of continuum components). The
code also provides for several different limb-darkening and
limb-brightening laws (see \citealt{Svoboda2009} for a discussion of
different limb-darkening laws).

Figure~\ref{fig:line_models} shows a comparison of the
\texttt{relline} model to models commonly used in X-ray astronomy. A
comparison of the model with an exact numerical evaluation of
Eq.~\ref{eq:lspec_general} that does not make use of precalculated
quantities and interpolation shows that there is no significant
deviation between both approaches (Fig.~\ref{fig:line_models}a). In
addition, for $a\ge 0$ the produced shape is very similar to the
\texttt{kyrline} model, which uses a table a factor 10 times larger
and a Gaussian emission profile instead of a delta function
(Fig.~\ref{fig:line_models}b). This result shows that for $a\ge 0$ and
when it is sufficient to use the limb-darkening law of
\citet{Laor1991} or the limb-brightening law of \citet{Haardt1993a},
both models can be used with confidence. For completeness,
Fig.~\ref{fig:line_models}c and Fig.~\ref{fig:line_models}d compare
the exact profile to the \texttt{kerrdisk} and the \texttt{laor}
model. Spikes in the former model are due to divergences in the
integration of the transfer function (see Sect.~\ref{sec:relline}). We
note, however, that if the line is evaluated on an energy grid
appropriate for a Silicon detector, these spikes will be averaged out
and therefore have no effect on any of the published results.
The \texttt{laor}-model, on the other hand, shows strong deviations
from the correct line shape which are caused by the coarse energy
grid. Especially in the tail of the line these deviations are large
enough that they could bias model fitting. For this reason, we caution
against using this model in data analysis work.

\section{Discussion and Conclusions}\label{sec:discussion}

\begin{figure}
    \includegraphics[width=84mm]{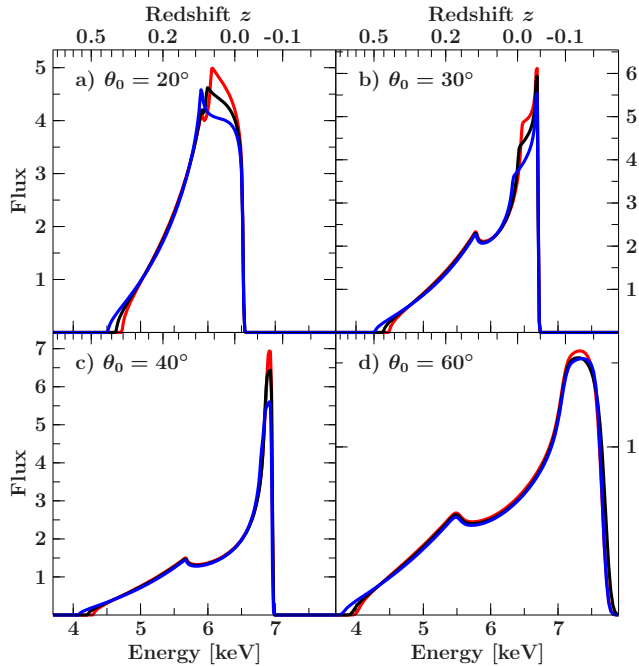}
    \caption{\label{fig:line_profiles} Line profiles of a relativistic
      iron line emitted at $6.4\,$keV in the rest-frame of the
      disc. Typical inclination angles $\theta_\mathrm{o}$ are
      displayed along with an emissivity of $r^{-3}$. The maximal
      spinning black hole ($a=+0.998$) is drawn in red (dashed in the
      printed version), the non-rotating ($a=0$) in black
      (dashed-dotted), and the blue (solid) line shows the broad
      emission line for maximal negative spin ($a=-0.998$). In order
      to allow comparison of the line shapes, the inner edge of the
      accretion disc was set to $r=9\,r_\mathrm{g}$ for all profiles.}
\end{figure}

\begin{figure}
    \includegraphics[width=84mm]{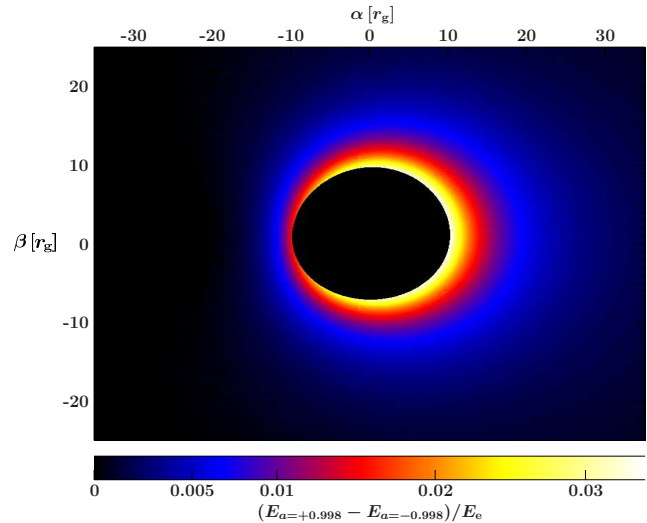}
    \caption{\label{fig:diff_image} Difference
        between the energy shifts experienced by photons from a
        maximally positively and a maximally negatively spinning
        accretion disc with inner radius
        $r_\mathrm{in} = 9\,r_\mathrm{g}$, viewed under an
        inclination of $40^\circ$.}
\end{figure}

In Figure~\ref{fig:line_profiles} we compare the line profiles for a
maximally rotating Kerr black hole, a Schwarzschild black hole, and a
black hole which is maximally counterrotating for several different
inclinations. The accretion disc emissivity was assumed to be
$\epsilon\propto r^{-3}$, i.e., the emissivity obtained from a simple
accretion disc in the Newtonian regime. This dependence is assumed in
many of the earlier models \citep[e.g.,][]{Fabian1989,Laor1991}. Note
that this assumption can be easily dropped in all modern line models.
In order to allow for a comparison of the line shapes with earlier
results, we use the limb-darkening law of \citet{Laor1991}, even
though for lines caused by fluorescence due to the irradiation of a
disc with hard X-rays from above, a limb-brightening law would be more
appropriate \citep{Svoboda2009}. In order to illustrate the pure
frame-dragging effects of these different spins onto the line shape,
the inner disc radius was set to $9\,r_\mathrm{g}$ for all three
spins. The Figure shows that in this case the major difference between
the different spins is the relative strength of the core of the line
to the red wing, which decreases with decreasing $a$. For this case of
a large inner radius, the most significant differences in line shape
are seen for low values of $\theta_\mathrm{o}$ while the red tails are
virtually indistinguishable. The slight increase in line flux at the
lowest energies is due to the increased Doppler boosting in the case
of $a<0$ (for a given radius, $u^t$ increases with decreasing $a$,
cf.~Eq.~\ref{eq:four_velocity}). The difference in energy
  shift of photons emerging from an accretion disc between maximal
  positive and maximal negative spin of the black hole is shown in
  Fig.~\ref{fig:diff_image}. As these differences are highest close to
  inner edge of the disc, a higher emissivity pronounces the devations
  in the line profiles. Moreover this figure shows that for an
  accretion disc with an inner radius larger than $30\,r_\mathrm{g}$
  no significant differences would be expected.

\begin{figure}
    \includegraphics[width=84mm]{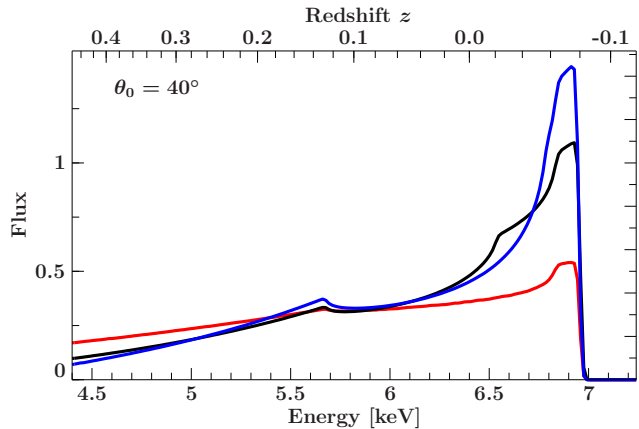}
    \caption{\label{fig:line_risco} A broadened emission line with
      $E_0 = 6.4\,\mathrm{keV}$ for different spins of the black hole,
      emerging from an accretion disc which is assumed to extend down
      to the marginal stable radius. Common parameters were chosen for
      the inclination angle ($\theta_\mathrm{o} = 40^\circ$) and the
      emissivity ($r^{-3}$). The color code is that of
      Fig.~\ref{fig:line_profiles}, i.e., $a=+0.998$ is drawn in red
      (dashed), $a=0$ in black (dashed-dotted), and $a=-0.998$ in blue
      (solid).}
\end{figure}

Figure~\ref{fig:line_risco} shows line profiles for different spins of
the black hole for the more realistic case that the disc extends down
to the marginally stable orbit. Since the inner edge of the disc is
closer to the black hole for positively spinning black holes,
more strongly redshifted photons emerge. As already noted by
  \citet{Jaroszynski1997a}, this leads to broader lines in these
  systems, especially for discs with an emissivity that is strongly
peaked towards $r_\mathrm{in}$. Maximally negatively spinning black
holes have the smallest width, although the line will still be
detectable as being broad even at CCD resolution (depending on
inclination, typical widths of the main peak are around 200\,eV).
Lines from counterrotating black holes will therefore be more
difficult to detect than lines from positively rotating black holes.

The major difference of line shapes for discs around black holes with
$a=0$ and counterrotating discs (see Fig.~\ref{fig:line_risco}) lies
in the strength of the blue peak, since the skew symmetric shape is
mainly due to frame dragging effects and the small inner radii.
Detecting these lines observationally is therefore more difficult than
detecting lines from discs around a positively rotating black hole. In
addition, as shown by \citet{Svoboda2009} limb-darkening/-brightening
affects the strength of the red wing. For counterrotating black holes,
this results in a possible degeneracy, as for different limb-darkening
laws similar line shapes might result for $a\sim 0$ and $a=-0.998$.
Using a physically motivated limb-darkening law would avoid this
degeneracy. The line shapes also become more similar if the assumption
that emission down to the radius of marginal stability contributes to
the shape is dropped. This assumption might not be justified in some
cases, as fluorescent emission only takes place in irradiated parts of
the disc which are not fully ionized. Thus the inner radius of the
emission becomes larger, which results in a weaker red tail of the
line profile. This effect leads to line shapes for different spin that
are more similar and closer to the ones in Fig.~\ref{fig:line_profiles}.

\begin{figure}
  \includegraphics[width=84mm]{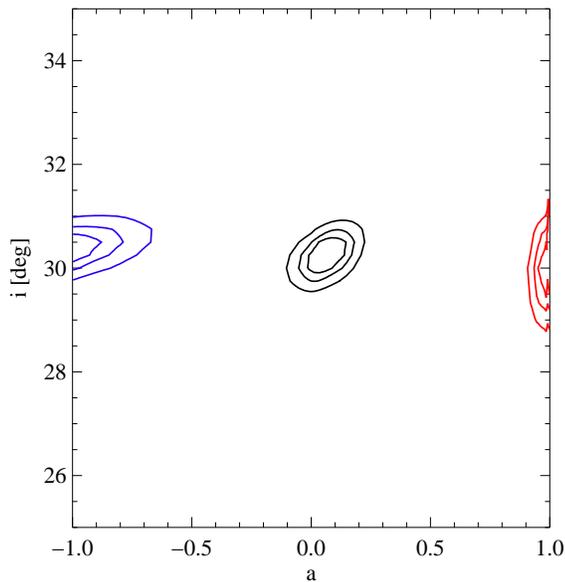}
  \caption{\label{fig:contour} Confidence contours for simulated
    50\,ksec observations of MCG$-$6-30-15 with \textsl{IXO}.
    Confidence contours at 68\%, 90\% and 99\% confidence are shown
    for three simulations assuming $a=-0.998$, $a=0$, and $a=+0.998$,
    respectively. In the determination of the contours, the continuum
    parameters ($N_\mathrm{H}$, $\Gamma$, and continuum normalization)
    and the line energy were left free.}
\end{figure}

  In order to study the question of observability in greater
  detail, we have performed simulations of observations of a
  relativistic line with the planned International X-ray Observatory
  (\textsl{IXO}), using response matrices obtained from the IXO team
  (Smith, priv.\ comm.). We base the simulations on power-law fits to
  \textsl{XMM-Newton} data from MCG$-$6-30-15 in a typical state,
  using a power law continuum with a 2--10\,keV flux of $2.5\times
  10^{-11}\,\mathrm{erg}\,\mathrm{cm}^{-2}\,\mathrm{s}^{-1}$ and
  photon index $\Gamma=1.6$, absorbed by a column
  $N_\mathrm{H}=10^{21}\,\mathrm{cm}^{-2}$. We set the equivalence
  width of the line to 350\,eV (typical for
  MCG$-$6-30-15). Figure~\ref{fig:contour} shows that in a 50\,ksec
  observation the next generation X-ray instrumentation will easily
  allow to separate even the difficult case of negatively spinning
  black holes.

In summary, motivated by predictions of negatively spinning black
holes from cosmology and from birth scenarios for Galactic black holes
\citep{brandt1995a,Volonteri2005a,king2008a}, we have presented new
results on the shape of relativistic lines from accretion discs around
black holes for the case that the angular momenta of the disc and the
black hole are aligned and counter-aligned \citep{King2005a}. We have
shown that lines from counterrotating discs are narrower than those
from Schwarzschild black holes, since the marginally stable orbit
moves outwards as the black hole's angular momentum decreases. Since
these lines still have a slight asymmetric shape, we still expect them
to be observable. We have also presented a new and more flexible
scheme for the calculation of line profiles for black holes of all
possible angular momenta and for arbitrary emissivity and
limb-darkening laws, which has a significantly smaller footprint in
terms of the amount of precalculation required. The model is available
from {\scriptsize
 \texttt{www.sternwarte.uni-erlangen.de/research/relline/}}.
Comparison of the model shows that for the case $a\ge 0$ its results
are in agreement with the modern relativistic line model of
\citet{Dovciak2004a}.

\emph{Acknowledgments.} We acknowledge partial support from the
European Commission under contract ITN 215212 ``Black Hole Universe''.
We thank John Davis for the development of the \texttt{SLxfig} module
used to prepare the figures in this paper and Manfred Hanke and Roland
Speith for useful comments. We thank the referee for insightful
comments that significantly improved the paper.

\bibliographystyle{mn2e} \bibliography{mnemonic,mn_abbrv,bib,agn,bhc,ns}

\begin{thebibliography}{}

\bibitem[\protect\citeauthoryear{{Arnaud}}{{Arnaud}}{1996}]{Arnaud1996a}
{Arnaud} K.~A.,  1996, in {G.~H.~Jacoby \& J.~Barnes} ed., Astronomical Data
  Analysis Software and Systems V Vol.~101 of Astronomical Society of the
  Pacific Conference Series, {XSPEC: The First Ten Years}.
pp 17--+

\bibitem[\protect\citeauthoryear{{Bardeen}, {Press} \& {Teukolsky}}{{Bardeen}
  et~al.}{1972}]{Bardeen1972}
{Bardeen} J.~M.,  {Press} W.~H.,    {Teukolsky} S.~A.,  1972, ApJ, 178, 347

\bibitem[\protect\citeauthoryear{Blum, Miller, Fabian, Miller, Homan, van~der
  Klis, Cackett \& Reis}{Blum et~al.}{2009}]{blum2009a}
Blum J.~L.,  Miller J.~M.,  Fabian A.~C.,  Miller M.~C.,  Homan J.,  van~der
  Klis M.,  Cackett E.~M.,    Reis R.~C.,  2009, ApJ, 706, 60

\bibitem[\protect\citeauthoryear{{Boyer} \& {Lindquist}}{{Boyer} \&
  {Lindquist}}{1967}]{Boyer1967}
{Boyer} R.~H.,  {Lindquist} R.~W.,  1967, Journal of Mathematical Physics, 8,
  265

\bibitem[\protect\citeauthoryear{{Brandt} \& {Podsiadlowski}}{{Brandt} \&
  {Podsiadlowski}}{1995}]{brandt1995a}
{Brandt} N.,  {Podsiadlowski} P.,  1995, MNRAS, 274, 461

\bibitem[\protect\citeauthoryear{{Brenneman} \& {Reynolds}}{{Brenneman} \&
  {Reynolds}}{2006}]{Brenneman2006a}
{Brenneman} L.~W.,  {Reynolds} C.~S.,  2006, ApJ, 652, 1028

\bibitem[\protect\citeauthoryear{{Caballero-Garc{\'i}a}, Miller, {D{\'i}az
  Trigo}, Kuulkers, Fabian, {Mas-Hesse}, Steeghs \& {van der
  Klis}}{{Caballero-Garc{\'i}a} et~al.}{2009}]{caballerogarcia2009a}
{Caballero-Garc{\'i}a} M.~D.,  Miller J.~M.,  {D{\'i}az Trigo} M.,  Kuulkers
  E.,  Fabian A.~C.,  {Mas-Hesse} J.~M.,  Steeghs D.,    {van der Klis} M.,
  2009, ApJ, 692, 1339

\bibitem[\protect\citeauthoryear{Cackett, Miller, Bhattacharyya, Grindlay,
  Homan, van~der Klis, Miller, Strohmayer \& Wijnands}{Cackett
  et~al.}{2008}]{cackett2008a}
Cackett E.~M.,  Miller J.~M.,  Bhattacharyya S.,  Grindlay J.~E.,  Homan J.,
  van~der Klis M.,  Miller M.~C.,  Strohmayer T.~E.,    Wijnands R.,  2008,
  ApJ, 674, 415

\bibitem[\protect\citeauthoryear{Cackett, Miller, Homan, van~der Klis, Lewin,
  {M\'endez}, Raymond, Steeghs \& Wijnands}{Cackett
  et~al.}{2009}]{cackett2009a}
Cackett E.~M.,  Miller J.~M.,  Homan J.,  van~der Klis M.,  Lewin W. H.~G.,
  {M\'endez} M.,  Raymond J.,  Steeghs D.,    Wijnands R.,  2009, ApJ, 690,
  1847

\bibitem[\protect\citeauthoryear{Carter}{Carter}{1968}]{Carter1968}
Carter B.,  1968, Phys. Rev., 174, 1559

\bibitem[\protect\citeauthoryear{Comastri, Brusa \& Civano}{Comastri
  et~al.}{2004}]{comastri:04a}
Comastri A.,  Brusa M.,    Civano F.,  2004, MNRAS, 351, L9

\bibitem[\protect\citeauthoryear{{Corral}, {Page}, {Carrera}, {Barcons},
  {Mateos}, {Ebrero}, {Krumpe}, {Schwope}, {Tedds} \& {Watson}}{{Corral}
  et~al.}{2008}]{corral:08a}
{Corral} A.,  {Page} M.~J.,  {Carrera} F.~J.,  {Barcons} X.,  {Mateos} S.,
  {Ebrero} J.,  {Krumpe} M.,  {Schwope} A.,  {Tedds} J.~A.,    {Watson} M.~G.,
  2008, A\&A, 492, 71

\bibitem[\protect\citeauthoryear{{Cunningham}}{{Cunningham}}{1975}]{Cunningham%
1975}
{Cunningham} C.~T.,  1975, ApJ, 202, 788

\bibitem[\protect\citeauthoryear{{Cunningham} \& {Bardeen}}{{Cunningham} \&
  {Bardeen}}{1973}]{Cunningham1973}
{Cunningham} J.~M.,  {Bardeen} C.~T.,  1973, ApJ, 183, 237

\bibitem[\protect\citeauthoryear{di Salvo, {D'A\'i}, Iaria, Burderi,
  Dov\v{c}iak, Karas, Matt, Papitto, Piraino, Riggio, Robba \&
  Santangelo}{di~Salvo et~al.}{2009}]{disalvo2009a}
di Salvo T.,  {D'A\'i} A.,  Iaria R.,  Burderi L.,  Dov\v{c}iak M.,  Karas V.,
  Matt G.,  Papitto A.,  Piraino S.,  Riggio A.,  Robba N.~R.,    Santangelo
  A.,  2009, MNRAS, 398, 2022

\bibitem[\protect\citeauthoryear{{Done} \& {Diaz Trigo}}{{Done} \& {Diaz
  Trigo}}{2010}]{done2010a}
{Done} C.,  {Diaz Trigo} M.,  2010, MNRAS, pp 1097--+

\bibitem[\protect\citeauthoryear{{Dov{\v c}iak}, {Karas} \& {Yaqoob}}{{Dov{\v
  c}iak} et~al.}{2004}]{Dovciak2004a}
{Dov{\v c}iak} M.,  {Karas} V.,    {Yaqoob} T.,  2004, Astrophysical Journal
  Supplement, 153, 205

\bibitem[\protect\citeauthoryear{{Fabian}, {Rees}, {Stella} \&
  {White}}{{Fabian} et~al.}{1989}]{Fabian1989}
{Fabian} A.~C.,  {Rees} M.~J.,  {Stella} L.,    {White} N.~E.,  1989, MNRAS,
  238, 729

\bibitem[\protect\citeauthoryear{{Fabian}, {Sanders}, {Ettori}, {Taylor},
  {Allen}, {Crawford}, {Iwasawa}, {Johnstone} \& {Ogle}}{{Fabian}
  et~al.}{2000}]{fabian2000a}
{Fabian} A.~C.,  {Sanders} J.~S.,  {Ettori} S.,  {Taylor} G.~B.,  {Allen}
  S.~W.,  {Crawford} C.~S.,  {Iwasawa} K.,  {Johnstone} R.~M.,    {Ogle} P.~M.,
   2000, MNRAS, 318, L65

\bibitem[\protect\citeauthoryear{{Fabian}, {Zoghbi}, {Ross}, {Uttley}, {Gallo},
  {Brandt}, {Blustin}, {Boller}, {Caballero-Garcia}, {Larsson}, {Miller},
  {Miniutti}, {Ponti}, {Reis}, {Reynolds}, {Tanaka} \& {Young}}{{Fabian}
  et~al.}{2009}]{fabian2009a}
{Fabian} A.~C.,  {Zoghbi} A.,  {Ross} R.~R.,  {Uttley} P.,  {Gallo} q.~C.,
  {Brandt} W.~N.,  {Blustin} A.~J.,  {Boller} T.,  {Caballero-Garcia} M.~D.,
  {Larsson} J.,  {Miller} J.~M.,  {Miniutti} G.,  {Ponti} G.,  {Reis} R.~C.,
  {Reynolds} C.~S.,  {Tanaka} Y.,    {Young} A.~J.,  2009, Nat, 459, 540

\bibitem[\protect\citeauthoryear{{Garofalo}}{{Garofalo}}{2009}]{Garofalo2009a}
{Garofalo} D.,  2009, ApJ, 699, 400

\bibitem[\protect\citeauthoryear{{Garofalo}, {Evans} \& {Sambruna}}{{Garofalo}
  et~al.}{2010}]{Garofalo2010a}
{Garofalo} D.,  {Evans} D.~A.,    {Sambruna} R.~M.,  2010, MNRAS, 406, 975

\bibitem[\protect\citeauthoryear{{Haardt}}{{Haardt}}{1993}]{Haardt1993a}
{Haardt} F.,  1993, ApJ, 413, 680

\bibitem[\protect\citeauthoryear{Houck \& Denicola}{Houck \&
  Denicola}{2000}]{Houck2000a}
Houck J.~C.,  Denicola L.~A.,  2000, in Manset N.,  Veillet C.,   Crabtree D.,
  eds, Astronomical Data Analysis Software and Systems IX No.~216 in ASP Conf.
  Ser., Isis: An interactive spectral interpretation system for high resolution
  x-ray spectroscopy.
p.~591

\bibitem[\protect\citeauthoryear{{Jaroszynski}}{{Jaroszynski}}{1997}]{Jaroszyn%
ski1997a}
{Jaroszynski} M.,  1997, Acta Astronomica, 47, 399

\bibitem[\protect\citeauthoryear{{Kataoka}, {Reeves}, {Iwasawa}, {Markowitz},
  {Mushotzky}, {Arimoto}, {Takahashi}, {Tsubuku}, {Ushio}, {Watanabe}, {Gallo},
  {Madejski}, {Terashima}, {Isobe}, {Tashiro} \& {Kohmura}}{{Kataoka}
  et~al.}{2007}]{Kataoka2007a}
{Kataoka} J.,  {Reeves} J.~N.,  {Iwasawa} K.,  {Markowitz} A.~G.,  {Mushotzky}
  R.~F.,  {Arimoto} M.,  {Takahashi} T.,  {Tsubuku} Y.,  {Ushio} M.,
  {Watanabe} S.,  {Gallo} L.~C.,  {Madejski} G.~M.,  {Terashima} Y.,  {Isobe}
  N.,  {Tashiro} M.~S.,    {Kohmura} T.,  2007, PASJ, 59, 279

\bibitem[\protect\citeauthoryear{Kerr}{Kerr}{1963}]{Kerr1963}
Kerr R.~P.,  1963, Phys. Rev. Lett., 11, 237

\bibitem[\protect\citeauthoryear{{King}, {Lubow}, {Ogilvie} \&
  {Pringle}}{{King} et~al.}{2005}]{King2005a}
{King} A.~R.,  {Lubow} S.~H.,  {Ogilvie} G.~I.,    {Pringle} J.~E.,  2005,
  MNRAS, 363, 49

\bibitem[\protect\citeauthoryear{{King}, {Pringle} \& {Hofmann}}{{King}
  et~al.}{2008}]{king2008a}
{King} A.~R.,  {Pringle} J.~E.,    {Hofmann} J.~A.,  2008, MNRAS, 385, 1621

\bibitem[\protect\citeauthoryear{{Krolik}}{{Krolik}}{1999}]{Krolik1999}
{Krolik} J.~H.,  1999, {Active galactic nuclei: from the central black hole to
  the galactic environment}.
Princeton Univ.\ Press, Princeton

\bibitem[\protect\citeauthoryear{{Laor}}{{Laor}}{1991}]{Laor1991}
{Laor} A.,  1991, ApJ, 376, 90

\bibitem[\protect\citeauthoryear{{Lindquist}}{{Lindquist}}{1966}]{Lindquist196%
6a}
{Lindquist} R.~W.,  1966, Annals of Physics, 37, 487

\bibitem[\protect\citeauthoryear{{Longinotti}, {de La Calle}, {Bianchi},
  {Guainazzi} \& {Dov{\v c}iak}}{{Longinotti} et~al.}{2008}]{Longinotti2008a}
{Longinotti} A.~L.,  {de La Calle} I.,  {Bianchi} S.,  {Guainazzi} M.,
  {Dov{\v c}iak} M.,  2008, Memorie della Societa Astronomica Italiana, 79, 259

\bibitem[\protect\citeauthoryear{{Martocchia}, {Matt}, {Karas}, {Belloni} \&
  {Feroci}}{{Martocchia} et~al.}{2002}]{Martocchia2002a}
{Martocchia} A.,  {Matt} G.,  {Karas} V.,  {Belloni} T.,    {Feroci} M.,  2002,
  A\&A, 387, 215

\bibitem[\protect\citeauthoryear{Miller, Fabian, Reynolds, Nowak, Homan,
  Freyberg, Ehle, Belloni, Wijnands, {van der Klis}, Charles \& Lewin}{Miller
  et~al.}{2004}]{miller2004a}
Miller J.~M.,  Fabian A.~C.,  Reynolds C.~S.,  Nowak M.~A.,  Homan J.,
  Freyberg M.~J.,  Ehle M.,  Belloni T.,  Wijnands R.,  {van der Klis} M.,
  Charles P.~A.,    Lewin W. H.~G.,  2004, ApJ, 606, L131

\bibitem[\protect\citeauthoryear{Miller, Fabian, Wijnands, Remillard,
  Wojdowski, Schulz, {Di Matteo}, Marshall, Canizares, Pooley \& Lewin}{Miller
  et~al.}{2002}]{miller2002a}
Miller J.~M.,  Fabian A.~C.,  Wijnands R.,  Remillard R.~A.,  Wojdowski P.,
  Schulz N.~S.,  {Di Matteo} T.,  Marshall H.~L.,  Canizares C.~R.,  Pooley D.,
     Lewin W. H.~G.,  2002, ApJ, 578, 348

\bibitem[\protect\citeauthoryear{Miller, Reynolds, Fabian, Cackett, Miniutti,
  Raymond, Steeghs, Reis \& Homan}{Miller et~al.}{2008}]{miller2008a}
Miller J.~M.,  Reynolds C.~S.,  Fabian A.~C.,  Cackett E.~M.,  Miniutti G.,
  Raymond J.,  Steeghs D.,  Reis R.,    Homan J.,  2008, ApJ, 679, L113

\bibitem[\protect\citeauthoryear{Miller, Reynolds, Fabian, Miniutti \&
  Gallo}{Miller et~al.}{2009}]{miller2009a}
Miller J.~M.,  Reynolds C.~S.,  Fabian A.~C.,  Miniutti G.,    Gallo L.~C.,
  2009, ApJ, 697, 900

\bibitem[\protect\citeauthoryear{{Miniutti}, {Fabian}, {Anabuki}, {Crummy},
  {Fukazawa}, {Gallo}, {Haba}, {Hayashida}, {Holt}, {Kunieda}
  et~al.,}{{Miniutti} et~al.}{2007}]{Miniutti2007amnras}
{Miniutti} G.,  {Fabian} A.~C.,  {Anabuki} N.,  {Crummy} J.,  {Fukazawa} Y.,
  {Gallo} L.,  {Haba} Y.,  {Hayashida} K.,  {Holt} S.,  {Kunieda} H.,
  et~al., 2007, PASJ, 59, 315

\bibitem[\protect\citeauthoryear{{Nandra}, {O'Neill}, {George} \&
  {Reeves}}{{Nandra} et~al.}{2007}]{nandra2007a}
{Nandra} K.,  {O'Neill} P.~M.,  {George} I.~M.,    {Reeves} J.~N.,  2007,
  MNRAS, 382, 194

\bibitem[\protect\citeauthoryear{Reynolds \& Fabian}{Reynolds \&
  Fabian}{2008}]{reynolds2008a}
Reynolds C.~S.,  Fabian A.~C.,  2008, ApJ, 675, 1048

\bibitem[\protect\citeauthoryear{{Reynolds}, {Garofalo} \&
  {Begelman}}{{Reynolds} et~al.}{2006}]{Reynolds2006a}
{Reynolds} C.~S.,  {Garofalo} D.,    {Begelman} M.~C.,  2006, ApJ, 651, 1023

\bibitem[\protect\citeauthoryear{Reynolds \& Nowak}{Reynolds \&
  Nowak}{2003}]{reynolds2003a}
Reynolds C.~S.,  Nowak M.~A.,  2003, Phys. Rep., 377, 389

\bibitem[\protect\citeauthoryear{Ross \& Fabian}{Ross \&
  Fabian}{2007}]{ross2007a}
Ross R.~R.,  Fabian A.~C.,  2007, MNRAS, 381, 1697

\bibitem[\protect\citeauthoryear{{Schnittman}}{{Schnittman}}{2006}]{Schnittman%
n2006a}
{Schnittman} J.~D.,  2006, ArXiv Astrophysics e-prints

\bibitem[\protect\citeauthoryear{Shaposhnikov, Titarchuk \&
  Laurent}{Shaposhnikov et~al.}{2009}]{shaposhnikov2009a}
Shaposhnikov N.,  Titarchuk L.,    Laurent P.,  2009, ApJ, 699, 1223

\bibitem[\protect\citeauthoryear{{Speith}, {Riffert} \& {Ruder}}{{Speith}
  et~al.}{1995}]{Speith1995}
{Speith} R.,  {Riffert} H.,    {Ruder} H.,  1995, Computer Physics
  Communications, 88, 109

\bibitem[\protect\citeauthoryear{Streblyanska, Hasinger, Finoguenov, Barcons,
  Mateos \& Fabian}{Streblyanska et~al.}{2005}]{streblyanska2005a}
Streblyanska A.,  Hasinger G.,  Finoguenov A.,  Barcons X.,  Mateos S.,
  Fabian A.~C.,  2005, A\&A, 432, 395

\bibitem[\protect\citeauthoryear{{Svoboda}, {Dov{\v c}iak}, {Goosmann} \&
  {Karas}}{{Svoboda} et~al.}{2009}]{Svoboda2009}
{Svoboda} J.,  {Dov{\v c}iak} M.,  {Goosmann} R.,    {Karas} V.,  2009, A\&A,
  507, 1

\bibitem[\protect\citeauthoryear{Tanaka, Nandra, Fabian, Inoue, Otani, Dotani,
  Hayashida, Iwasawa, Kii, Kunieda, Makino \& Matsuoka}{Tanaka
  et~al.}{1995}]{tanaka1995a}
Tanaka Y.,  Nandra K.,  Fabian A.~C.,  Inoue H.,  Otani C.,  Dotani T.,
  Hayashida K.,  Iwasawa K.,  Kii T.,  Kunieda H.,  Makino F.,    Matsuoka M.,
  1995, Nat, 375, 659

\bibitem[\protect\citeauthoryear{{Thorne}}{{Thorne}}{1974}]{1974Thorne}
{Thorne} K.~S.,  1974, ApJ, 191, 507

\bibitem[\protect\citeauthoryear{{Volonteri}, {Madau}, {Quataert} \&
  {Rees}}{{Volonteri} et~al.}{2005}]{Volonteri2005a}
{Volonteri} M.,  {Madau} P.,  {Quataert} E.,    {Rees} M.~J.,  2005, ApJ, 620,
  69

\bibitem[\protect\citeauthoryear{Wilms, Reynolds, Begelman, Reeves, Molendi,
  Staubert \& Kendziorra}{Wilms et~al.}{2001}]{wilms2001a}
Wilms J.,  Reynolds C.~S.,  Begelman M.~C.,  Reeves J.,  Molendi S.,  Staubert
  R.,    Kendziorra E.,  2001, MNRAS, 328, L27

\bibitem[\protect\citeauthoryear{{Wise}, {McNamara}, {Nulsen}, {Houck} \&
  {David}}{{Wise} et~al.}{2007}]{wise2007a}
{Wise} M.~W.,  {McNamara} B.~R.,  {Nulsen} P.~E.~J.,  {Houck} J.~C.,    {David}
  L.~P.,  2007, ApJ, 659, 1153

\bibitem[\protect\citeauthoryear{Yamada, Makishima, Uehara, Nakazawa,
  Takahashi, Dotani, Ueda, Ebisawa, Kubota \& Gandhi}{Yamada
  et~al.}{2009}]{yamada2009a}
Yamada S.,  Makishima K.,  Uehara Y.,  Nakazawa K.,  Takahashi H.,  Dotani T.,
  Ueda Y.,  Ebisawa K.,  Kubota A.,    Gandhi P.,  2009, ApJ, 707, L109

\end{thebibliography}

\appendix
\label{lastpage}

\end{document}